\documentclass[preprintnumbers,amsmath,amssymb,aps]{revtex4}

\usepackage{graphicx}
\usepackage{multirow}
\usepackage{tabularx}

\newcommand{\bee}{\begin{equation}}
\newcommand{\ee}{\end{equation}}
\newcommand{\beea}{\begin{eqnarray}}
\newcommand{\eea}{\end{eqnarray}}

\begin{document}

\title{Reweighting QCD simulations with dynamical overlap fermions}
\author{Thomas DeGrand}
\affiliation{
Department of Physics, University of Colorado,
Boulder, CO 80309 USA
}

\begin{abstract}  I apply a recently developed algorithm for reweighting
simulations of lattice QCD from one quark mass to another to simulations
performed with overlap fermions in the epsilon regime. I test it by computing the condensate
from distributions of the low lying eigenvalues of the Dirac operator.
Results seem favorable.
\end{abstract}
\maketitle

\section{Recalling the Algorithm}

Several recent papers\cite{Hasenfratz:2008fg,LP} have described the
 idea of performing simulations of QCD at one value of the quark mass by reweighting
the configurations generated in a simulation at another quark mass.
Overlap fermions\cite{Neuberger:1997fp,Neuberger:1998my}, 
the most easy to analyze and and most expensive
to simulate of
all lattice discretizations of the Dirac operator, are natural candidates for reweighting.
I describe some rather trivial numerical experiments I have made, reweighting 
simulations with two flavors of overlap fermions. The techniques are basically those of
A.~Hasenfratz, R.~Hoffmann and S.~Schaefer \cite{Hasenfratz:2008fg},
 but the properties of the overlap operator
eliminate many technical difficulties.

Let's summarize the implementation of reweighting I will follow, by considering a single
flavor of fermion with Dirac operator $D(m)$. We will not worry about questions of
positivity or reality for the moment.
Let's suppose that we have generated a stream of configurations
 with one value of the quark mass $m_1$
and want, by reweighting, to use the stream to perform 
simulations at another quark mass $m_2$.
To do so, one must reweight configurations by a factor
\bee
w = \frac{\det D(m_2)}{\det D(m_1)}  .
\label{eq:w}
\ee
Introducing the operator
\bee
\Omega=D(m_2)^{-1}D(m_1)
\ee
we can write $w$ as an average over a set of complex random vectors $\xi$,
\bee
w = \frac{\int D\xi \exp(-\xi^\dagger \Omega \xi)}{\int D\xi \exp(-\xi^\dagger \xi )}
\equiv \langle \exp(-\xi^\dagger(\Omega-1)\xi )\rangle_\xi   .
\label{eq:w2}
\ee

With the usual definition of the link gauge variable $U$ and pure gauge action $S_g$,
expectation values of operators evaluated at mass $m_2$ are given by
\beea
\langle O\rangle_{m_2} &=& \frac{1}{Z_2} \int dU e^{-S_g} \det D(m_2) \  O(U) \nonumber \\
 &=& \frac{1}{Z_2} \int dU e^{-S_g} \det D(m_1) ({\det}^{-1}\Omega) O(U) \nonumber \\
 &=& \frac{Z_1}{Z_2}\langle O(U)\exp(-\xi^\dagger(\Omega-1)\xi)\rangle_{m_1,\xi} 
\nonumber \\
\eea
where 
\bee
\frac{Z_2}{Z_1} = \langle \exp(-\xi^\dagger(\Omega-1)\xi) \rangle_{m_1,\xi} .
\ee
If we imagine that our simulation at mass $m_1$ consists of a stream of pairs of variables
$\{U_i,\xi_i\}$, then the expectation value is
\bee
\langle O(U)\rangle_{m_2} = \frac{\sum_i O(U_i) \exp(-\xi_i^\dagger(\Omega(U_i)-1)\xi_i) }
{\sum_i \exp(-\xi_i^\dagger(\Omega(U_i)-1)\xi_i )}.
\ee
That is, we reweight each configuration by a factor
\bee
w_i = \exp(-\xi_i^\dagger(\Omega(U_i)-1)\xi_i).
\label{eq:sfactor}
\ee

This was so far all quite general.   Now we
assume that we are doing simulations with overlap fermions. For any number of flavors,
all calculations can be performed using the
squared Hermitian Dirac operator 
$D(m_i)^\dagger D(m_i)= H(m_i^2)=s_i H(0)^2 + m_i^2$ where $s_i=1- m_i^2/(4R_0)^2$ and $H(0)$
is the massless squared overlap Hermitian Dirac operator
\bee
H(0)=r_0(\gamma_5 - \epsilon(h)).
\ee
The quantity $h$ is the kernel operator $h=\gamma_5(d-R_0)$ (in terms of a kernel Dirac operator $d$)
and $\epsilon(h)$ is the matrix sign 
function. We assume that we have recorded
a set of eigenfunctions of $H(0)$ (and their associated eigenvalues),
 $H(0)|k\rangle = \lambda_k |k\rangle$. The spectrum of $H(0)^2$ consists of
a set of zero eigenvalue chiral modes and a set of degenerate (paired) nonzero eigenvalue
eigenmodes of opposite parity. The nonzero mode contribution to $w$  in Eq. \ref{eq:w2}
can be computed using random vectors $\xi$ which are chiral, with chirality in the sector 
without zero modes. Each flavor of dynamical fermion  has its own chiral random vector.

Now we come to the question of practicality:
Reweighting will fail if the weight of each configuration deviates widely from the mean,
because then only the (presumably small number) of configurations carrying a large weight will 
contribute to averages. It can also fail if the estimator (Eq. \ref{eq:sfactor}) has a large variance,
for then one will need to average  the same underlying gauge configuration over many estimators.
Can schemes be devised, so that the weights $w_i$
do not fluctuate too much from configuration to configuration?
Presumably what will work will depend on the simulation and reweighted quark mass and the 
simulation volume. The phase space of possible choices is large.

It is always  a good thing to replace as much of  the stochastic estimator of the determinant
with an exact result. Introducing the Hermitian projector onto low eigenmodes of $H(0)^2$
(call it $P$ and its complement $\bar P = 1 - P$),
\bee
\det \Omega= \det P \Omega \ \det \bar P \Omega.
\ee
We compute $P\Omega$ from eigenvalues
and we only need to make a stochastic estimator for the high eigenmode part of the weight $w_i$
\bee
w_i = (\frac{m_2}{m_1})^{N_0} \prod_{k=N_0+1}^N \frac{s_2 \lambda_k^2 + m_2^2}{s_1 \lambda_k^2 + m_1^2} 
\exp(-\xi_i^\dagger(\bar P \Omega \bar P - 1)\xi_i) \equiv w_{low} w_{high}
\ee
(reweighting a configuration with $N_0$ zero modes from mass $m_1$ to $m_2$, 
and considering a single flavor).
To complete the equation set,
\bee
\Omega = \frac{s_2}{s_1} + \frac{c_{12}}{H(m_2)^2}
\ee
where $c_{12}=m_1^2 - (s_1/s_2)m_2^2$,
and with $y=\bar P \xi$ because the random vector can only live in the space of $\bar P \Omega \bar P$,
\bee
w_{high}= \exp(-(y^\dagger (\Omega -1) y ).
\ee
Projection of low modes plus the use of a random vector in the chirality space without zero modes
 improves the effective conditioning number of $H(m_2)^{-2}$.

One interesting place in parameter space is the epsilon regime. Here the quarks are so light that
the pion ``fills the box'' -- if the volume is $V=L^4$, then $m_\pi L <<1$ (and all other
mass scales $M$ large, $ML>1$) defines the epsilon regime.
Let's perform some experiments there: I have several sample data sets with
 $N_f=2$ flavors of overlap fermions on $12^4$ simulation volumes at
a nominal lattice spacing of 0.14 fm. I will use a quark mass $am_q=0.03$ fm
(nominally about 43 MeV) and $am_q=0.01$. They were generated using the hybrid Monte Carlo algorithm,
with the reflection/refraction algorithm devised in Ref.~\cite{Fodor:2003bh}. They used
the differentiable hypercubic smeared link of
 Ref. \cite{Hasenfratz:2007rf} and one or two additional heavy
pseudo-fermion fields as suggested by Hasenbusch\cite{Hasenbusch:2001ne}.
The integration is done with multiple-time scales\cite{Urbach:2005ji}.
 Details of the actions are described in
 Refs.
\cite{DeGrand:2000tf,DeGrand:2004nq,DeGrand:2006ws,DeGrand:2007tm,DeGrand:2006nv}.
All in all, these are very conventional overlap fermions.
 I typically compute the lowest 12 eigenvectors and eigenvalues of
$H(0)^2$; these eigenvalues run up to about $\lambda \sim 0.04$.

\begin{figure}
\begin{center}
\includegraphics[width=0.6\textwidth,clip]{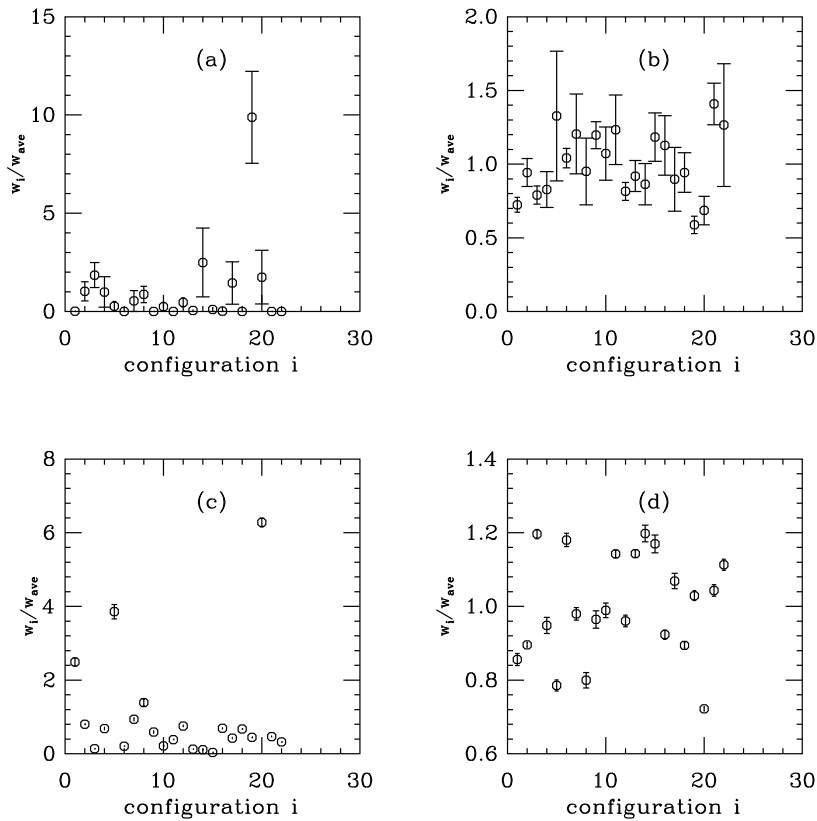}
\end{center}
\caption{Determinant weight from a set of $am_q=0.03$ configurations into (a) $am_q=0.01$,
(b) $am_q=0.035$, normalized by the average weight, computed
without removing eigenmodes. Panels (c) and (d) show the reweighting to mass
$am_q=0.01$ and 0.035 while
treating the 12 lowest eigenvalues of $H(0)$ exactly.
\label{fig:comp03}
}
\end{figure}

The first test checks what can gained by removing eigenmodes. We take a set of
 22 lattices from our stream of $am_q=0.03$ simulations and reweight them to a set of target masses:
$am_q=0.01$ and  0.035. In this test we averaged the stochastic part of the weight
over six pairs of two chiral pseudofermions.   Fig. \ref{fig:comp03} 
show a comparison of the resulting weights either not keeping any eigenmodes or removing the lowest
12 eigenmodes from the stochastic estimator. 
The error bars show the variation in weight over the ensemble of pseudofermion noise 
vectors used for each configuration.
 It is clear that the choice of removing eigenmodes is the superior one, from the point of view of suppressing
the variance  of the estimator.

\begin{figure}
\begin{center}
\includegraphics[width=0.6\textwidth,clip]{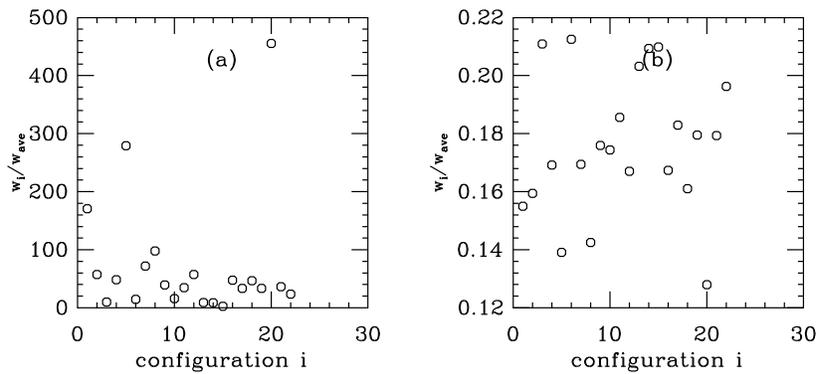}
\end{center}
\caption{Contribution to determinant weight from the 12 lowest eigenvalues of $H(0)^2$ for the
$am_q=0.03$ configurations of Fig. \protect\ref{fig:comp03}: (a) $am_q=0.01$, (b) $am_q=0.035$.
\label{fig:individ03ll}
}
\end{figure}

The low eigenvalues do not capture the entire reweighting factor. Fig.~\ref{fig:individ03ll}
shows the weights from just the low eigenvalues (divided by the average reweighting factor from the
true weights). Their
(incorrectly normalized) values appear to track to full weight.

\section{Eigenmode distributions in the epsilon regime}
By themselves, pictures of the fluctuating weights give no indication of how well
an actual reweighted calculation will perform. A test is needed.
For a little physics example I select the problem of determining the condensate from the 
distribution
of low-lying eigenvalues of the massless Dirac operator in the epsilon regime in
sectors of fixed topology. These distributions are given by 
Random Matrix Theory (RMT)
\cite{Shuryak:1992pi,Verbaarschot:1993pm,Verbaarschot:1994qf,Damgaard:2000qt,Damgaard:2000ah}.
Overlap fermions are optimal for this project (as for any epsilon regime simulation) due to the
control they give over lattice topology.

I am aware of three previous measurements of $\Sigma_L$ from eigenvalues
 with $N_f=2$ flavors of dynamical
overlap fermions. Two of them,
Refs. \cite{DeGrand:2006nv} and \cite{DeGrand:2007tm}.
were not really in the epsilon regime; in the second paper, the bare quark 
mass is $am_q=0.03$
corresponding to a pion mass in lattice units of $am_\pi=0.324$ (so $m_\pi L \sim 3.9$).
The JLQCD collaboration,  Ref.\cite{Fukaya:2007yv}, has a true epsilon regime calculation of $\Sigma$. 
I am also aware of two recent studies which use dynamical fermions which are not exactly chiral,
 but which are said to have highly improved chiral symmetry: Refs \cite{Lang:2006ab} 
and \cite{Hasenfratz:2007yj}. (The latter simulation used 2+1 flavors.)
All of these papers produce similar and unsurprising
values for the condensate, $\Sigma \sim (250 \ {\rm MeV} )^3$.
An unpleasant feature of the epsilon regime is that finite volume corrections are power law,
not exponential.
The effect is to replace the value of the condensate extracted from the RMT fit,
$\Sigma$, by $\Sigma_L = \rho_\Sigma \Sigma$
where
\bee
\rho_\Sigma = 1 + \frac{N_f^2-1}{N_f} \frac{1}{F^2} \Delta(0) +\dots
\ee
with $\Delta(0)$  the contribution to the tadpole graph (propagator
at zero separation) from finite-volume image terms. In the epsilon
regime, $\Delta(0) = -\beta_0/\sqrt{V}$
and $\beta_0$ depends on the geometry\cite{Gasser:1986vb}. (It is 0.1405 for
hypercubes.) 

I carry two data sets into -- or closer to -- the epsilon regime.
The first data set is the set of $am_q=0.03$ configurations from  Ref. \cite{DeGrand:2007tm},
a stream of 400 thermalized hybrid Monte Carlo trajectories of unit length
giving 30 $\nu=0$ lattices and 75 $|\nu|=1$ ones. The second set is a collection of 74 
$\nu=0$ lattices from four little streams (each of about 150 trajectories) run 
at $am_q=0.01$. This is already on the edge of the epsilon regime. Nearly all were $\nu=0$.
I reweighted the $am_q=0.03$ data set to target masses $am_q=0.01$, 0.015, 0.02, 0.025, and 0.035.
I reweighted the $am_q=0.01$ set to 0.0025, 0.005, 0.0075, 0.025 and 0.03.
I used only a single pair of pseudofermions per configuration, since I treated 
the lowest 12 eigenvalues
of $H(0)^2$ exactly.

\begin{figure}
\begin{center}
\includegraphics[width=0.6\textwidth,clip]{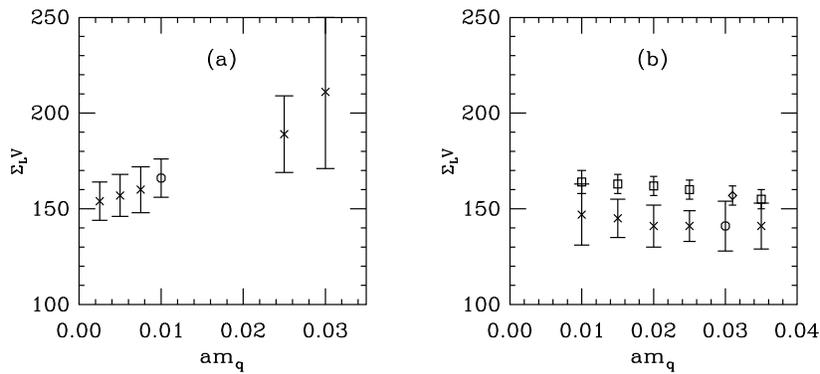}
\end{center}
\caption{The quantity $\Sigma_L V$ from reweighted $12^4$ lattices, as a function of
the target reweighted mass. (a) True mass $am_q=0.01$, $\nu=0$. Crosses are reweighted points,
the octagon is unweighted. (b) True mass $am_q=0.03$: $\nu=0$ lattices labeled
as in (a), $\nu=1$ lattices
are marked with squares when reweighted, diamond when not.
\label{fig:rewtsigma}
}
\end{figure}

The data were analyzed with a conventional bootstrap analysis. In the bootstrap, the weight
 of a configuration was the number of times it was selected for the bootstrap, times the
 (normalized) weight factor from the determinant ratio. Results for $\Sigma_L V$ from a fit to
 the lowest eigenvalue distribution in one topological sector are shown in 
Fig. \ref{fig:rewtsigma}. Of course, all the reweighted points at different quark masses
are highly correlated; they came from the same data sets. It appears that
 reweighting into the epsilon regime
was successful, while trying to go to larger quark masses ($am_q=0.01$ to 0.035, for example)
was less so. It is probably no surprise that reweighting for a small change in
 mass works better than reweighting a big change.

Readers might recall that to complete a calculation of $\Sigma_L$, one needs a separate
determination of a lattice spacing and a lattice-to-continuum matching factor $Z_S$.
$Z_S$ was determined for this action in Ref. \cite{DeGrand:2007tm}:
$Z_S^{\overline{MS}}$(2~GeV)=0.76(3).
Of course, the lattice spacing varies as the bare parameters of the simulation change.
However, this variation is small in the epsilon regime simply because the absolute change in the 
quark mass is small. For this data set, $r_0/a= 3.71(5)$ at $am_q=0.03$ and 3.77(7) at $am_q=0.01$.
Thus the three unreweighted values of $r_0^3 \Sigma_L$ from this study are
0.326(30) ($am_q=0.01$, $\nu=0$), 0.347(37) ($am_q=0.03$, $\nu=0)$, and 0.294(20) ($am_q=0.03$, $|\nu|=1$,
and with $r_0 \sim 0.5$ fm, $\Sigma_L \sim (260-270$ MeV)${}^3$.

\section{Conclusions}
Reweighting dynamical overlap fermion data sets into the epsilon regime worked better than I expected.
Groups doing simulations with overlap fermions might well be advised to investigate it as a technique.
All the ingredients will probably already be in hand.
The main reason for reweighting nonchiral actions -- 
namely, that one wants to avoid exceptional
configurations -- obviously does not apply to overlap fermions. However, overlap fermion simulations
are so expensive that running at many parameter values is daunting. Any methodology which allows
one to recycle old configurations is worth exploiting.

\section*{Acknowledgments}

I would like to thank A.~Hasenfratz and S. Schaefer for discussions and correspondence.
This work was supported in part by the US Department of Energy.

\end{document}